\documentclass{article}

\usepackage{arxiv}

\usepackage{graphicx} % Required for inserting images
\usepackage{cite}
\usepackage{subcaption}
\usepackage{amssymb}
\usepackage{amsfonts}
\usepackage{amsmath}
\usepackage{longtable}
\usepackage{multirow}
\usepackage{adjustbox}

\title{SMEFT operators in rare multi-top processes}

\newif\ifuniqueAffiliation
\ifuniqueAffiliation % Standard variant of author block
\author{ {\hspace{1mm}A.~Aleshko} \\
	\And
	{\hspace{1mm}E~.Boos} \\
	Skobeltsyn Institute of Nuclear Physics\\
	Lomonosov Moscow State University\\
	119991 Moscow, Russia \\
	\texttt{boos@theory.sinp.msu.ru} \\
}
\else
\usepackage{authblk}

\author[1]{%
	A.~Aleshko\thanks{\texttt{isserq@gmail.com}}%
}
\author[1]{%
	E~.Boos\thanks{\texttt{boos@theory.sinp.msu.ru}}%
}
\author[1]{%
	V.~Bunichev\thanks{\texttt{bunichev@theory.sinp.msu.ru}}%
}
\author[1]{%
	L.~Dudko\thanks{\texttt{dudko@sinp.msu.ru}}%
}
\affil[1]{Skobeltsyn Institute of Nuclear Physics, Lomonosov Moscow State University, 119991 Moscow, Russia}
\fi

\renewcommand{\shorttitle}{SMEFT operators in rare multi-top processes}

\begin{document}

\maketitle

\begin{abstract}
\noindent
Nowadays, the Standard Model Effective Field Theory (SMEFT) provides a standard framework to parameterize potential deviations from the Standard Model and to combine information from multiple processes in global analyses. This review summarizes dedicated studies that constrain dimension-six Wilson coefficients using three top-quark and four top-quark production processes. We highlight the complementarity of these channels, as well as summarize the main problems and prospects in the area. A concise introduction to the SMEFT formalism and a discussion of the problem of potential perturbative unitarity violation are also provided.
\end{abstract}

\section{Introduction}

The top-quark occupies a privileged position in collider searches for physics beyond the Standard Model (BSM). In many UV-motivated scenarios (e.g. supersymmetry, composite Higgs/top-partner, top-philic dark sectors, etc.), the leading indirect effects preferentially modify top interactions and multi-top final states. This is reflected in the availability of a substantial body of works interpreting top-quark signatures in terms of concrete BSM scenarios and simplified models \cite{Arina_2016, Matsedonskyi_2016, Garny_2018, franceschini_2023, blasi_2024, atlas_2025_ttbar}. Consequently, processes with multiple top quarks in the final state are of prime interest for constraining potential effects of New Physics.

On the one hand, there are single-top and top-pair production, which constitute the precision backbone of the top program. For instance, ATLAS reports an inclusive $pp\xrightarrow{}t\bar{t}$ cross section at $\sqrt{s}=13$ TeV of $\sigma_{t\bar{t}}=829$ pb with a $\sim 2\%$ total uncertainty \cite{Aad_2023}, while CMS published $\sigma_{t\bar{t}}=791$ pb with a $\sim 3\%$ total uncertainty \cite{Tumasyan_2021}. Both measurements are consistent with state of the art predictions $\sigma_{t\bar{t}}^{NNLO+NNLL} = 832^{+55}_{-54}$ pb \cite{Aad_2023, Czakon_2013, Czakon_2014}. Similarly, the \textit{t}-channel single-top production cross-section is measured with several percent precision at $\sqrt{s}=13$, e.g. $\sigma_{tq +\bar{t}q}=221^{+13}_{-13}$ pb \cite{Aad_2024} (ATLAS) and $\sigma_{tq +\bar{t}q}=207^{+33}_{-33}$ \cite{Sirunyan_2020} (CMS). The corresponding SM predictions are available at NNLO QCD for inclusive rates (e.g. $\sigma(tq)=134.2^{+2.2}_{-2.2}$ pb and $\sigma(\bar{t}q)=80.0^{+1.6}_{-1.6}$ pb \cite{Aad_2024, Campbell_2021}), with residual scale and PDF uncertainties at the few percent level. A substantial literature exists on SMEFT interpretations of top-quark measurements, including dedicated global analyses of top-sector observables and broader multi-process global fits that incorporate top data \cite{Buckley_2015, Hartland_2019, Brivio_2020, ethier_2021, Ellis_2021}.

\begin{figure}[t!]
\centering
\includegraphics[width=0.24\textwidth,clip]{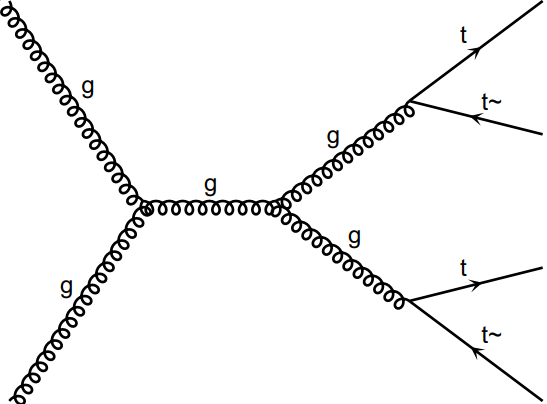}
\includegraphics[width=0.24\textwidth,clip]{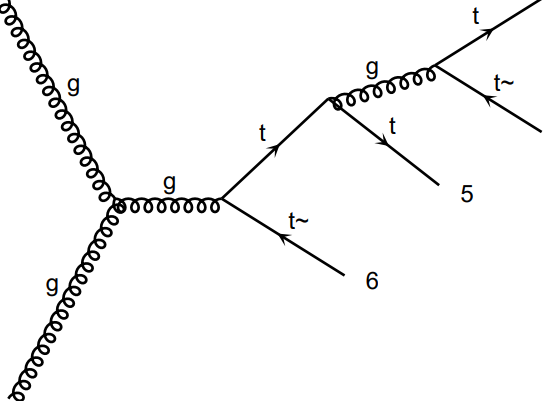}
\includegraphics[width=0.24\textwidth,clip]{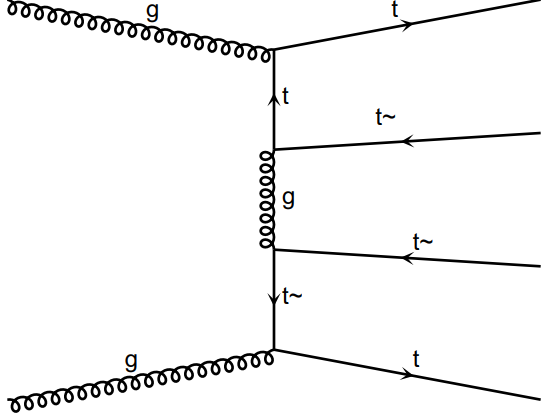}
\includegraphics[width=0.24\textwidth,clip]{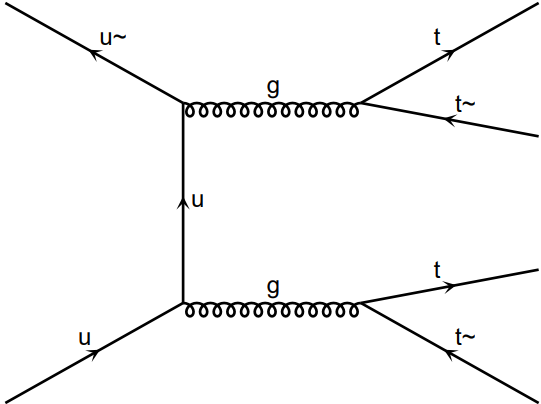}
\caption{Representative Feynman diagrams of four top quarks LO production in SM.}
\label{4topsm_diag}
\end{figure}

On the other hand, there are rare multi-top processes, such as three and four top-quark production, which complement this precision program in a qualitatively different way. A common motivation is that BSM effects can yield large relative enhancements, given the small SM rates, and that the operator structures can lead to enhanced EFT sensitivity, including strong energy dependence and potentially sizable contribution from formally subleading quadratic terms \cite{Zhang_2018, Banelli_2021, Blekman_2022, Aoude_2022}. Four-top production $pp\xrightarrow{}t\bar{t}t\bar{t}$ has now been observed by both CMS and ATLAS. CMS measures $\sigma_{t\bar{t}t\bar{t}}=17.7^{+4.4}_{-4.0}$ fb at $\sqrt{s}=13$ TeV \cite{Hayrapetyan_2023}, while ATLAS reports $\sigma_{t\bar{t}t\bar{t}}=22.5^{+6.6}_{-5.5}$ fb \cite{ATLAS:2023ajo}. From the theory side, four top-quark production was thoroughly studied at NLO \cite{Bevilacqua:2012em, Frederix:2017wme}, with the best available prediction given by NLO matched with threshold resummation: $\sigma_{t\bar{t}t\bar{t}}^{NLO+NLL^\prime} = 14.65^{+8.2\%}_{-17.4\%}$ \cite{PhysRevLett.131.211901}. Representative diagrams for the four top-quark SM production at Leading Order are given in Figure \ref{4topsm_diag}.
% Reference for the four top-quark production SM predictions is given in Table \ref{4top_sm_cs}.

% \begin{table}[t]
%     \resizebox{0.9\textwidth}{!}{%
%     %\begin{tabular}{p{0.3\textwidth} p{0.25\textwidth} p{0.15\textwidth} p{0.15\textwidth} p{0.15\textwidth}}
%     \begin{tabular}{l|c|c|c}
%     \multicolumn{4}{c}{4 top-quark production, LO} \\
%     \hline
%        C.o.m. energy, TeV & SM cros.-sect. $\sigma_{SM}$, fb & scale uncert., \% & PDF uncert., \% \\  \hline
%        13 & 9.02 & 73.4 &	7.32 \\
%        14 & 12.0 & 72.6 &	7.28 \\
%     \hline
%     \multicolumn{4}{c}{4 top production, NLO (EW + QCD)} \\
%     \hline
%        13 & 13.2 & 20.1 &	2.5 \\
%        14 & 17.8 & 20.3 &	2.5 \\
%     \end{tabular}
%     }
%     \caption{SM cross-sections for $pp \rightarrow t \bar{t} t \bar{t}$ process, given factorization/renormalization scale set to mass of the top-quark $\mu_{F/R}=m_t$ \cite{Aleshko:2023rkv}.}
%     \label{4top_sm_cs}
% \end{table}

\begin{figure}[t!]
\centering
\includegraphics[width=0.24\textwidth,clip]{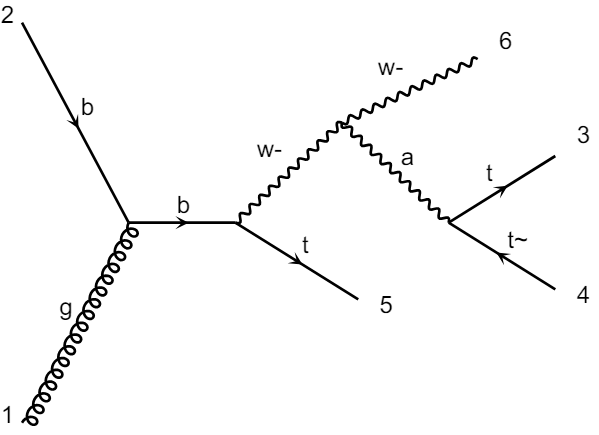}
\includegraphics[width=0.24\textwidth,clip]{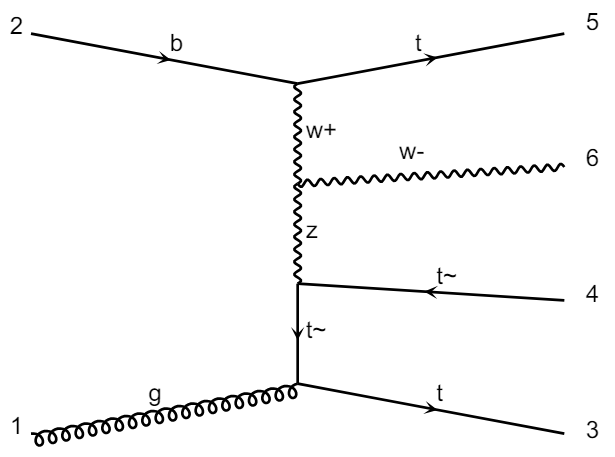}
\includegraphics[width=0.24\textwidth,clip]{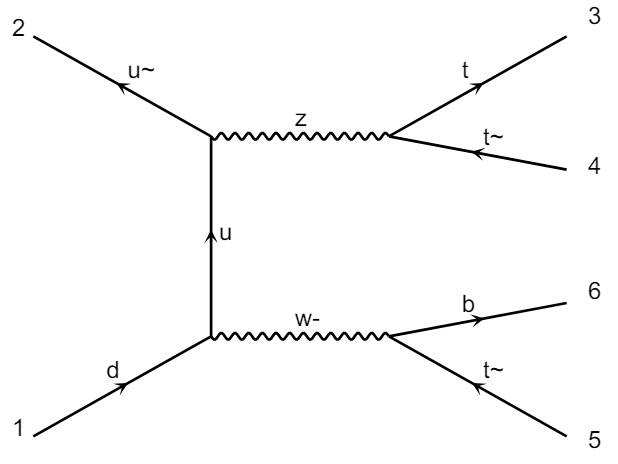}
\includegraphics[width=0.24\textwidth,clip]{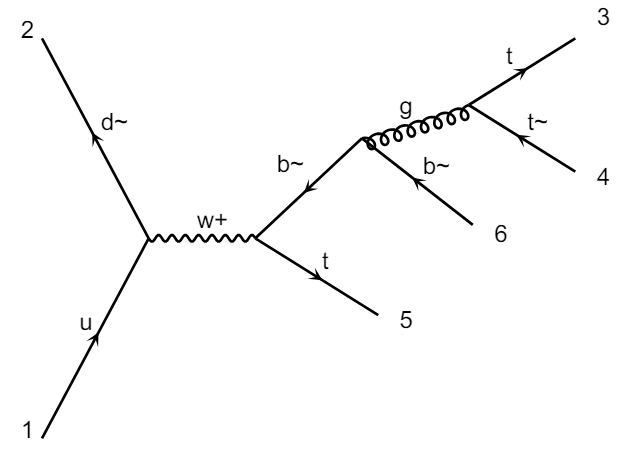}
\caption{Representative Feynman diagrams of three top quarks LO production in SM}
\label{3topsm_diag}
\end{figure}

Three-top production has not yet been observed, primarily because its SM rate is even smaller and its signatures strongly overlap with those of four top-quark and other multi-lepton backgrounds (notably $t\bar{t}W, t\bar{t}Z$ and misidentified leptons). In practice, three-top is often treated as an irreducible background in four top-quark analyses, and simultaneous fits exhibit strong anti-correlations between the processes, making a clean separation challenging \cite{Hayrapetyan_2023, ATLAS:2023ajo, shooshtari2026searchphysicsstandardmodel}. The SM prediction for the inclusive three top-quark production cross section is $\sigma_{t\bar{t}t(\bar{t})}\approx 1-2$ fb \cite{Boos:2021yat, Barger:2010uw}, with scale and PDF uncertainties at the level of $\sim30\%$ at LO. Representative tree-level diagrams for $pp \rightarrow t \bar{t} t(\bar{t}) + X$  are provided in Figure \ref{3topsm_diag}, and reference SM predictions are given in Table \ref{3top_sm_cs}. From a SMEFT perspective, three top-quark production is naturally sensitive to four-fermion contact operators, as well as to operators modifying top electroweak interactions, because it necessarily involves electroweak vertices, therefore potentially offering complementary directions in Wilson Coefficient space.

\begin{table}[t]
    \resizebox{0.9\textwidth}{!}{
    \begin{tabular}{l|c|c|c}
    \multicolumn{4}{c}{3 top-quark production, LO} \\
    \hline
       C.o.m. energy, TeV & SM cros.-sect. $\sigma_{SM}$, fb & scale uncert., \% & PDF uncert., \% \\  \hline
       13 & 1.16 & 30.1 &	7.1 \\
       14 & 1.5 & 29.2 &	6.64 \\
       
    \hline
    \multicolumn{4}{c}{3 top production, NLO (QCD)} \\
    \hline
       13 & 1.78 & 20.0 &	3.3 \\
       14 & 2.3 & 19.4 &	3.1 \\
    \end{tabular}
    }
    \caption{SM cross-sections for $pp \rightarrow t \bar{t} t(\bar{t}) + X$ process, given factorization/renormalization scale set to mass of the top-quark $\mu_{F/R}=m_t$ \cite{Aleshko:2023rkv}.}
    \label{3top_sm_cs}
\end{table}

The purpose of this review is to summarize the current status of constraints on dimension-six SMEFT operators probed by three and four top-quark production. The paper is organized as follows. Section \ref{sec:smeft_intro} introduces the SMEFT framework and fixes notation and assumptions. Section \ref{sec:34top} reviews the relevant multi-top processes, summarizes representative results from the literature, and highlights the main experimental and theoretical limitations. Section \ref{sec:unitary_contraints} discusses the EFT validity issue, with an emphasis on practical criteria based on perturbative (partial-wave) unitarity and the resulting implications for Wilson Coefficient constraints.

\section{SMEFT methodology}
\label{sec:smeft_intro}

The Standard Model Effective Field Theory (SMEFT) provides a systematic, model-independent parametrization of heavy New Physics in terms of higher-dimensional local operators built from SM fields and invariant under $SU(3)_c\times SU(2)_L\times U(1)_Y$ \cite{Buchmuller:1985jz, Degrande:2012wf, Aguilar-Saavedra:2018ksv, Brivio_2019, Boos:2022cys, Isidori_2024}. Two assumptions are at the core of SMEFT: the absence of new light particles and the linear realization of electroweak symmetry breaking. The former is justified by the fact that any new light particles should be very weakly coupled to the SM, while the latter is addressed within the framework of Higgs Effective Field Theory (HEFT) \cite{Brivio_2016, Gr_f_2023}. The operator content also depends on the assumed flavor structure: common choices range from the fully general case to Minimal Flavor Violation (MFV), which reduces the dimensionality of the Wilson-coefficient space. 

The SMEFT Lagrangian can be written schematically as:
\begin{equation}
    L = L_{SM} + \sum^{}_{n=5} \sum^{}_{k}{\frac{c^{\{n\}}_k}{\Lambda^{n-4}}O^{\{n\}}_k},
\label{eft_lagr}
\end{equation}

\noindent where $L_{SM}$ is the renormalizable SM Lagrangian, $\Lambda$ denotes the hypothetical scale of BSM physics, $O^{\{n\}}_i$ are gauge-invariant operators of canonical dimension $n$, and $c^{\{n\}}_i$ are dimensionless effective couplings, known as Wilson Coefficients (WC).

At dimension five there is a unique term - the Weinberg operator $O^{\{5\}}\sim HHll$ \cite{Weinberg:1979}, which violates lepton number and generates Majorana neutrino masses after electroweak symmetry breaking. Some collider probes specifically targeting lepton number violation can be interpreted in terms of the Weinberg operator or its UV completions (e.g. same-sign dilepton signatures mediated by heavy Majorana states \cite{Tumasyan_2023}). However, for collider observables in the top sector that conserve lepton number, this operator is typically irrelevant. Therefore, under the common assumption of baryon and lepton number conservation, dimension six operators provide the leading SMEFT contribution.

A general operator list contains redundancies: operators related by integration by parts, equations of motion/field redefinitions, Fierz and Bianchi identities are physically equivalent and should not be counted independently. Using these to eliminate related operators yields a non-redundant operator basis. For baryon number conserving dimension six SMEFT, the community standard is the Warsaw basis \cite{Grzadkowski:2010es}. Complete bases at dimension eight are also available \cite{Murphy_2020, Li_2021}. All dimension 7 \cite{Lehman_2014} and dimension 9 \cite{Li_2021_9} operators violate either lepton or baryon number and, therefore, are also generally omitted from consideration. Systematic counting and classification of operators was done up to dimension 12 \cite{henning201928430993}.

Phenomenological SMEFT results are commonly reported as confidence intervals on a chosen subset of Wilson Coefficients. The mapping between experimental results and EFT contributions is obtained by expanding each observable around its SM prediction:
\begin{equation}
    O_{exp} = O_{SM} + \sum^{}_{k}{\frac{c_k^{\{6\}}}{\Lambda^2}\sigma^{(1,6)}_k} + \sum^{}_{j <= k}{\frac{c_k^{\{6\}} c_j^{\{6\}}}{\Lambda^4}\sigma^{(2, 6)}_{k,j}} + \sum^{}_{k}{\frac{c_k^{\{8\}}}{\Lambda^4}\sigma^{(1,8)}_k} + O(\Lambda^{-6}),
\label{sigma_eq}
\end{equation}

\noindent where $O_{exp}$ - measured value for an observable, $O_{SM}$ is the SM prediction of one, $\sigma^{(1,n)}_k$ and $\sigma^{(2, n)}_{k,j}$ - numerical coefficients representing linear and quadratic (in terms of EFT coupling) contributions of effective operators. Values of $O_{SM}$ and $\sigma^{(i,d)}_k$ are obtained from perturbative calculations and/or Monte Carlo simulations.

Given experimental measurements $O_{exp}$ and calculated coefficients $\sigma^{(i,d)}_k$ with corresponding uncertainties, one then specifies a statistical procedure. A common choice is a multivariate Gaussian (equivalently a chi-squared) approximation using the full covariance matrix, while more differential or event-level analyses typically rely on Poisson likelihoods with nuisance parameters for experimental and theoretical systematics. Inference is then performed either in a frequentist framework (profile likelihood, confidence intervals) or a Bayesian framework (posterior sampling), depending on the analysis design and the available likelihood information.

A number of public tools are available for facilitating these steps. For LHC predictions in the Warsaw basis, SMEFTsim \cite{Brivio_2017} and SMEFTatNLO \cite{Degrande_2021} provide FeynRules/UFO implementation enabling automated event generation. For running and matching, widely used packages include DsixTools \cite{Celis_2017} and the Python library wilson \cite{Aebischer_2018}. For global or semi-global inference, frameworks such as EFTfitter \cite{Castro_2016}, HEPfit \cite{de_Blas_2020}, SMEFiT \cite{Giani_2023}, and smelli \cite{stangl2020smellismeftlikelihood} provide statistical engines and curated observable sets. Broader tool surveys are collected in community reports \cite{editors2019computingtoolssmeft, Aebischer_2024}.

Theoretical consistency can also restrict the admissible values of Wilson Coefficients. In particular, the requirement of perturbative (partial-wave) unitarity at high energies provides process dependent bounds; this aspect is discussed in Section \ref{sec:unitary_contraints}. More general constraints follow from analyticity and unitarity of the $S$-matrix and can be formulated as positivity bounds on certain combinations of Wilson Coefficients; for these, we refer to the dedicated literature \cite{Adams_2006, Remmen_2019, Yamashita_2021, Bellazzini_2021, Chala_2022, chala_2024}.

Finally, the end goal is often to translate SMEFT constraints into limits on parameters of a specific UV completion. This requires matching the UV model to a SMEFT basis. Detailed discussions and practical matching techniques are available in the literature \cite{henning2016oneloopmatchingrunningcovariant, Bakshi_2019, Fuentes_Mart_n_2023, Fuentes_Mart_n_2024}. All these steps may seem cumbersome, but ultimately it is the convenience of having parameterized model-independent deviations and a common language for combining heterogeneous datasets that has made SMEFT the leading tool for BSM analyses.

\begingroup

\setlength{\tabcolsep}{10pt} % Default value: 6pt
\renewcommand{\arraystretch}{1.2} % Default value: 1

\begin{table}[t!]
    % \resizebox{\textwidth}{!}{%
    %\begin{tabular}{p{0.3\textwidth} p{0.25\textwidth} p{0.15\textwidth} p{0.15\textwidth} p{0.15\textwidth}}
    \centering
    \begin{tabular}{l|l}
    \multicolumn{1}{c}{4H} & \multicolumn{1}{c}{2HB} \\
        \hline
    $O_{tt}^{1} = (\bar{t}_R\gamma^{\mu}t_R)(\bar{t}_R\gamma_{\mu}t_R)$ & $O_{tW} = i(\bar{Q}\sigma^{\mu \nu}\tau_{I}t_R)\widetilde{\varphi}W^I_{\mu \nu}$  + h.c.\\
    $O_{QQ}^{1} = (\bar{Q}_L\gamma^{\mu}Q_L)(\bar{Q}_L\gamma_{\mu}Q_L)$ & $O_{tB} = i(\bar{Q}\sigma^{\mu \nu}t_R)\widetilde{\varphi}B_{\mu \nu}$ + h.c.\\
    $O_{QQ}^{8} = (\bar{Q}_L\gamma^{\mu}T^{A}Q_L)(\bar{Q}_L\gamma_{\mu}T^{A}Q_L)$ & $O_{tG} =  i(\bar{Q}\sigma^{\mu\nu}T_{A}t)\widetilde{\varphi}G_{\mu\nu}^{A}$ + h.c.\\
    $O_{Qt}^{1} =(\bar{Q}_L\gamma^{\mu}Q_L)(\bar{t}_R\gamma_{\mu}t_R)$ & $O_{\varphi Q}^{-} = i(H^{\dagger}\stackrel{\leftrightarrow}{D}\varphi)(\bar{Q} \gamma^{\mu}Q) - i(\varphi^{\dagger}\stackrel{\leftrightarrow}{D^I}\varphi)(\bar{Q}\tau^I \gamma^{\mu}Q)$\\
    $O_{Qt}^{8} = (\bar{Q}_L\gamma^{\mu}T^{A}Q_L)(\bar{t}_R\gamma_{\mu}T^{A}t_R)$ & $O_{\varphi Q}^{-} = i(H^{\dagger}\stackrel{\leftrightarrow}{D}\varphi)(\bar{Q} \gamma^{\mu}Q) - i(\varphi^{\dagger}\stackrel{\leftrightarrow}{D^I}\varphi)(\bar{Q}\tau^I \gamma^{\mu}Q)$\\
    ~ & $O_{\varphi t} = i(\varphi^{\dagger}\stackrel{\leftrightarrow}{D}\varphi)(\bar{t}_R\gamma^{\mu}t_R)$ \\
    ~ & $O_{tp} = (\varphi^{\dagger}\varphi - v/2)\bar{Q}t\tilde{\varphi}$ + h.c. \\
    \hline
    \multicolumn{2}{c}{2H2L} \\
    \hline
    \multicolumn{2}{l}{$O_{Qq}^{1,1} = (\bar{q}\gamma^{\mu}q)(\bar{Q}\gamma_{\mu}Q) + \frac{1}{6}(\bar{Q}\gamma^{\mu}q)(\bar{q}\gamma_{\mu}Q) + \frac{1}{2}(\bar{q}\gamma^{\mu}Q)(\bar{Q}\gamma_{\mu}q)$} \\
    \multicolumn{2}{l}{$O_{Qq}^{1,3} =(\bar{q}\gamma^{\mu}\tau^I q)(\bar{Q}\gamma_{\mu}\tau^IQ) + \frac{1}{6}(\bar{Q}\gamma^{\mu} q)(\bar{q}\gamma_{\mu} Q) - \frac{1}{6}(\bar{q}\gamma^{\mu}\tau^I Q)(\bar{Q}\gamma_{\mu}\tau^I q)$} \\
    \multicolumn{2}{l}{$O_{Qq}^{1,8} = (\bar{Q}\gamma^{\mu}q)(\bar{q}\gamma_{\mu}Q) + 3(\bar{q}\gamma^{\mu}Q)(\bar{Q}\gamma_{\mu}q)$} \\
    \multicolumn{2}{l}{$O_{Qq}^{3,8} = (\bar{Q}\gamma^{\mu}q)(\bar{q}\gamma_{\mu}Q) - (\bar{q}\gamma^{\mu}Q)(\bar{Q}\gamma_{\mu}q)$} \\
    \multicolumn{2}{l}{$O_{tq}^{1} = (\bar{q}\gamma^{\mu}q)(\bar{t}_R\gamma_{\mu}t_R)$} \\
    \multicolumn{2}{l}{$O_{tq}^{8} = (\bar{q}\gamma^{\mu}T^A q)(\bar{t}_R\gamma_{\mu}T^At_R)$} \\
    \multicolumn{2}{l}{$O_{tu}^{1} = (\bar{u}_R\gamma^{\mu}u_R)(\bar{t}_R\gamma_{\mu}t_R) + \frac{1}{3}(\bar{u}_R\gamma^{\mu}t_R)(\bar{t}_R\gamma_{\mu}u_R)$} \\
    \multicolumn{2}{l}{$O^{8}_{tu} = 2(\bar{u}_R\gamma^{\mu}t_R)(\bar{t}_R\gamma_{\mu}u_R)$} \\
    \multicolumn{2}{l}{$O^{1}_{Qu} = (\bar{Q}\gamma^{\mu}Q)(\bar{u}_R\gamma^{\mu}u_R)$} \\
    \multicolumn{2}{l}{$O^{8}_{Qu} = (\bar{Q}\gamma^{\mu}T^A Q)(\bar{u}_R\gamma^{\mu}T^A u_R)$} \\
    \multicolumn{2}{l}{$O^{1}_{td} = (\bar{t}_R\gamma_{\mu}t_R)(\bar{d}_R\gamma^{\mu}d_R)$} \\
    \multicolumn{2}{l}{$O^{8}_{td} = (\bar{t}_R\gamma_{\mu}T^A t_R)(\bar{d}_R\gamma^{\mu}T^A d_R)$} \\
    \multicolumn{2}{l}{$O^{1}_{Qd} = (\bar{Q}\gamma^{\mu}Q)(\bar{d}_R\gamma^{\mu}d_R)$} \\
    \multicolumn{2}{l}{$O^{8}_{Qd} = (\bar{Q}\gamma^{\mu}T^A Q)(\bar{d}_R\gamma^{\mu}T^A d_R)$} \\
    \end{tabular}
    % }
    \vspace{5mm}
    \caption{Definitions of combinations of dimension 6 SMEFT operators, relevant for the top-sector. Combinations are defined to be consistent with the implementation in \cite{Degrande_2021}. In the upper left section are operators with 4 heavy fermions (4H), in the upper right one operators with 2 heavy and 2 light fermions (2H2L) are collected, and the lower one contains operators with 2 heavy fermions and boson fields (2HB).}
    \label{oper_def}
\end{table}

\endgroup

\section{Rare multi-top processes in SMEFT}
\label{sec:34top}

\subsection{Four top-quark production}

\begin{figure}[t]
    \centering
    \includegraphics[width=0.35\textwidth,clip]{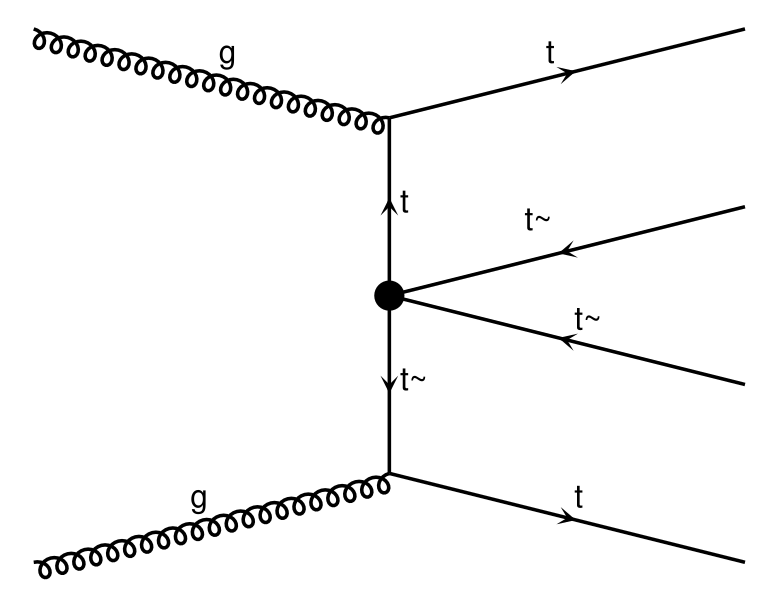}
    \includegraphics[width=0.35\textwidth,clip]{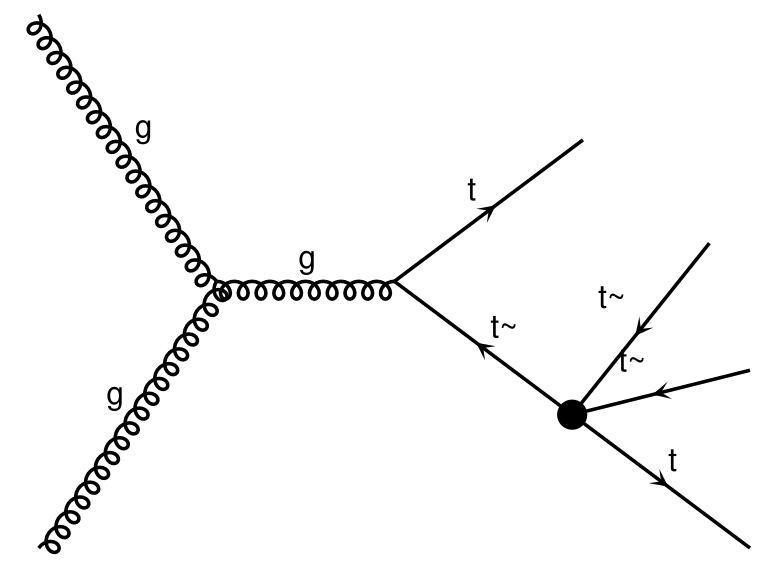}
\caption{Example diagrams for $pp \rightarrow t \bar{t} t\bar{t}$ with insertion of effective vertices, induced by four-heavy fermion SMEFT operators.}
\label{4top_diag_eft}
\end{figure}

Consider a four top-quark hadroproduction process in the presence of a New Physics (NP). In a general UV scenario $pp\xrightarrow{}t\bar{t}t\bar{t}$ can receive contributions due to heavy NP states. At leading EFT order all such possible effects are parameterized in the form of dimension 6 effective operators. Currently, the only available observable is the inclusive cross-section measured in multi-lepton final states, with $\sim 30\%$ uncertainties \cite{Hayrapetyan_2023, ATLAS:2023ajo}. In the absence of precise experimental data, a standard phenomenological strategy is to treat the SM prediction as a pseudo-measurement with conservative theory and systematic uncertainties. Such theoretical constraints are widely used to quantify prospective reach and to assess the impact of adding differential observables once they become available.

In SMEFT, four top-quark production is a particularly direct probe of four-fermion contact interactions involving top fields (4H section of Table \ref{oper_def}), and many analyses focus on this particular subset. Relevant representative diagrams are provided in Figure \ref{4top_diag_eft}. A number of works provide constraints on the corresponding dimension six operator subsets \cite{Zhang_2018, Hayrapetyan_2023, ATLAS:2023ajo, Aoude_2022, Degrande:2024mbg, Aleshko:2023aes, Aleshko:2023rkv}. The process also receives contributions from 2-Heavy-2-Light type of operators (2H2L section of Table \ref{oper_def}) through the $q\bar{q}$ initiated subchannel, as well as from two-fermion dipole and purely bosonic operators \cite{Aoude_2022, Degrande:2024mbg}. In practice, many phenomenological studies prioritize the four-heavy subset because (i) several 2H2L directions are already tightly constrained by $t\bar{t}$ and associated production, and (ii) the $gg$ initiated channel dominates the $q\bar{q}$ one as the collision energy increases, making this subset less important in projections to high-energy scenarios.

\begin{figure}[t]
\centering
\begin{subfigure}[tp]{0.48\textwidth}
    \centering
    \includegraphics[width=0.98\textwidth,clip]{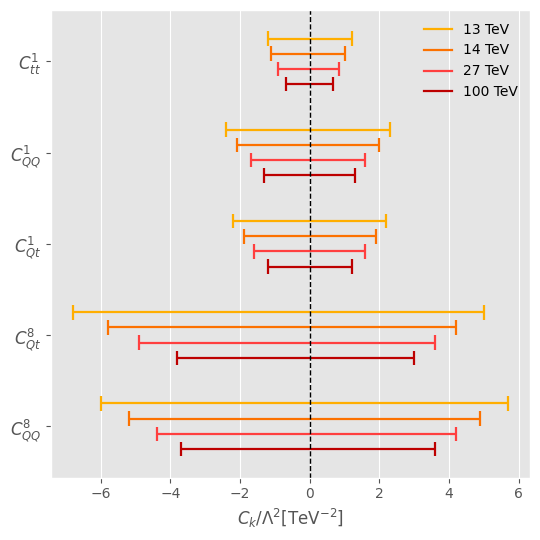}
    \caption{ }
\end{subfigure}
\begin{subfigure}[tp]{0.48\textwidth}
    \centering
    \includegraphics[width=0.98\textwidth,clip]{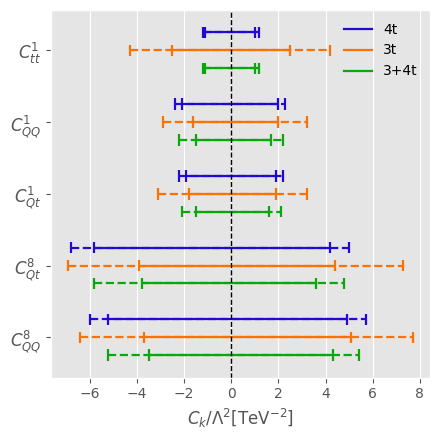}
    \caption{ }
\end{subfigure}
\vspace{-2mm}
\caption{(a) Projections to several high-energy scenarios of the expected limits on $C_k/\Lambda^2 \rm [TeV^{-2}]$ of 4-heavy fermion operators obtainable from four top-quark production process; (b) Comparison of the expected limits on $C_k/\Lambda^2 \rm [TeV^{-2}]$ estimated for $pp \rightarrow t\bar{t}t\bar{t}$ (4t) and $pp \rightarrow t\bar{t}t(\bar{t}) + X$ (3t) at $\sqrt{s}$ = 13 TeV (dotted) and $\sqrt{s}$ = 14 TeV (solid)~\cite{Aleshko:2023rkv}.}
% \label{smeft-dim6-4top}
\label{34t_comparison}
\end{figure}

Across the studies cited above, the extracted constraints on the relevant four-fermion WC are mutually consistent. In particular, bounds obtained from the $pp \rightarrow t \bar{t} t\bar{t}$ observation are in good correspondence with "theoretical" projections based on SM predictions supplemented with conservative uncertainty estimates. An illustrative example of such limits on the WC of four-heavy fermion operators is provided in Figure \ref{34t_comparison} \textbf{(a)}.

Several important points can be highlighted across the corresponding studies. First, the $t \bar{t} t\bar{t}$ production can already yield competitive constraints on four-fermion directions due to enhanced EFT sensitivity. Second, the attainable accuracy of present experimental fits is strongly limited by both experimental and theoretical uncertainties. Third, in prospective high-energy or high-luminosity scenarios, the potential improvement of the corresponding bounds is less dramatic than naively expected. Even when experimental uncertainties are assumed to decrease substantially, theory uncertainties remain dominant for this channel, effectively limiting the attainable precision. This implies that meaningful future gains in Wilson Coefficient constraints from four top-quark production require parallel progress in both experimental analyses and theoretical predictions, including improved perturbative accuracy and a controlled treatment of EFT validity/truncation effects.

Overall, the main challenges for precision SMEFT constraints from four top-quark production process are: (i) the limited number of statistically powerful observables, and (ii) theoretical uncertainties in both SM and SMEFT predictions. Hopes in the community are that progress along these lines would elevate four-top production from a single-number constraint to a genuinely discriminating probe in global analyses of BSM physics.

\subsection{Three top-quark production}

Within the Standard Model (SM), three top-quark production $pp \rightarrow t \bar{t} t(\bar{t}) + X$
is a very rare process. In contrast to four top-quark production, three-top final states arise only in association with additional objects, and are commonly decomposed into subprocesses such as $pp \rightarrow t \bar{t}tW$ and $pp \rightarrow t \bar{t}tj$, reflecting the electroweak origin of the process. To date, three-top production has not been observed as a standalone signal at the LHC. However, it is constrained indirectly in multi-lepton analyses designed for $t\bar{t}t\bar{t}$. In particular, the ATLAS four-top observation analysis reports 95\% CL intervals for the inclusive three top-quark production cross-section and for the $t\bar{t}tW$ and $t \bar{t}tj$ components \cite{ATLAS:2023ajo}. Similar to four top-quark production, predictions for three top-quark production also suffer from large theoretical uncertainties at LO, while also having a problem with an unambiguous definition of the process at NLO due to the presence of resonant diagrams in the real part of the correction.

\begin{figure}[t!]
% \centering
\begin{subfigure}{\textwidth}
\includegraphics[width=0.24\textwidth,clip]{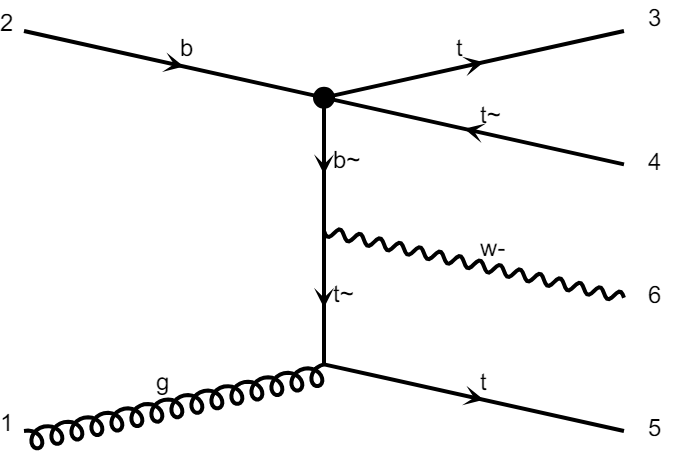}
\includegraphics[width=0.24\textwidth,clip]{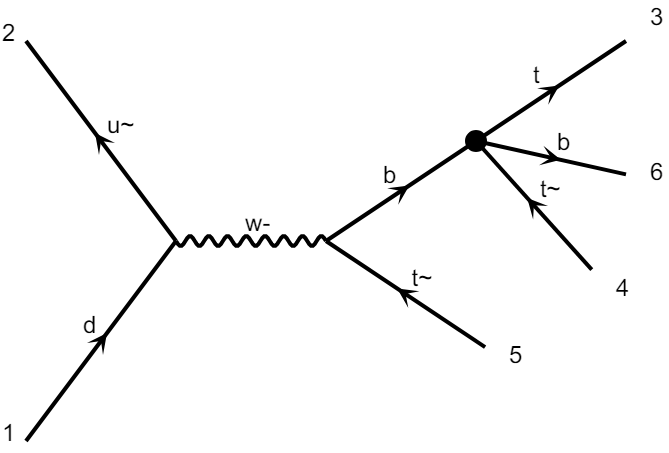}
\includegraphics[width=0.24\textwidth,clip]{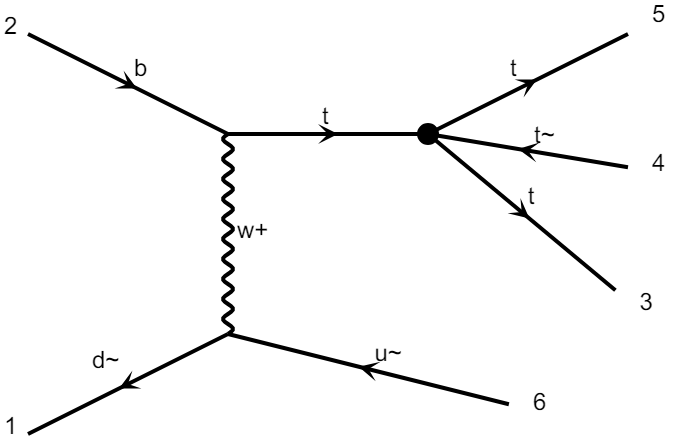}
\includegraphics[width=0.24\textwidth,clip]{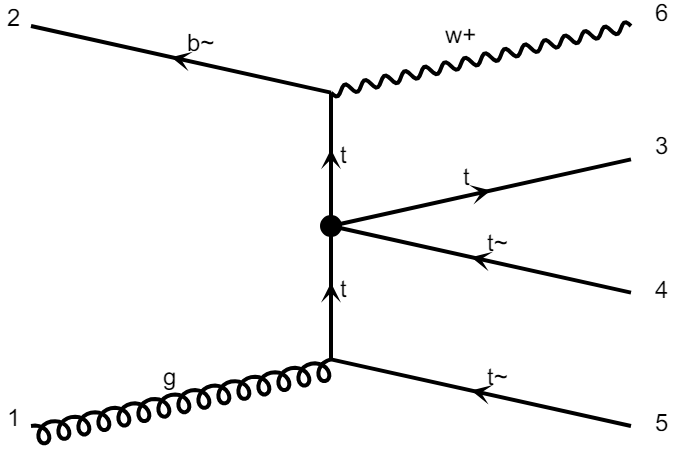}
\caption{4 heavy fermions}
\label{diag_eft:4h}
 \end{subfigure}

\begin{subfigure}{\textwidth}
\includegraphics[width=0.24\textwidth,clip]{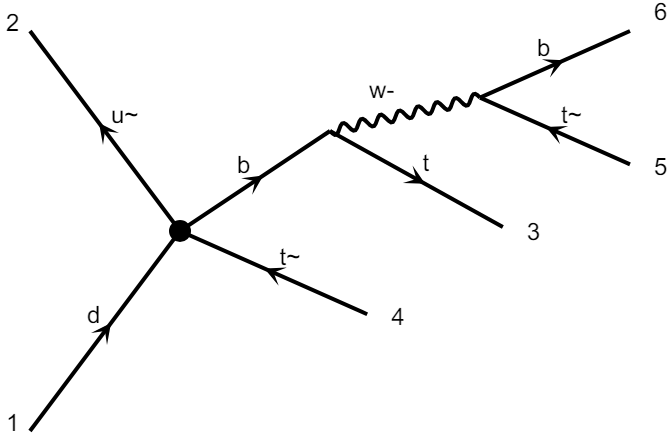}
\includegraphics[width=0.24\textwidth,clip]{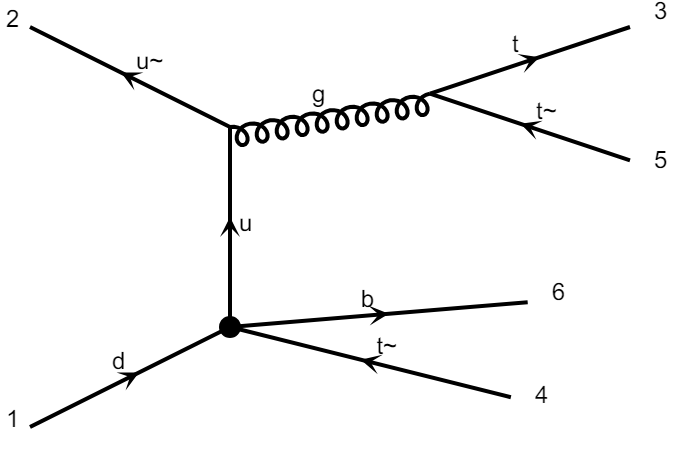}
\includegraphics[width=0.24\textwidth,clip]{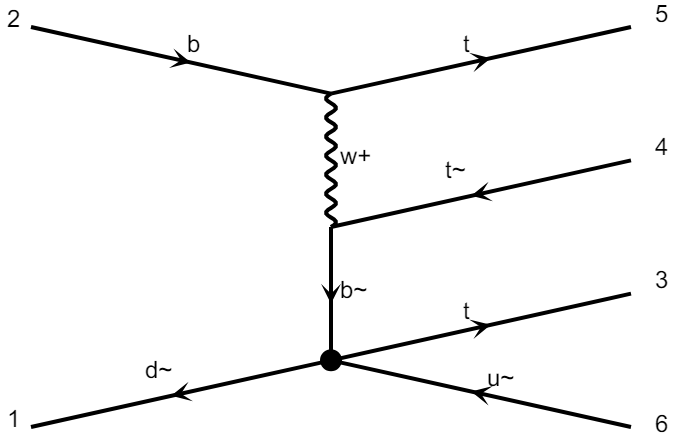}
\includegraphics[width=0.24\textwidth,clip]{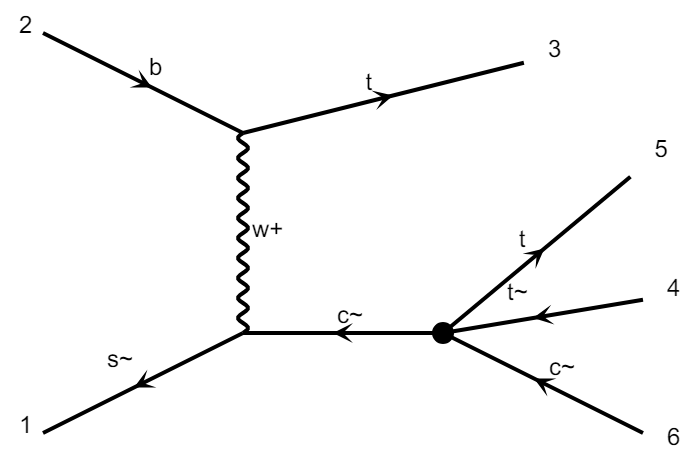}
\caption{2 heavy and 2 light fermions}
\label{diag_eft:2h2l}
 \end{subfigure}

\begin{subfigure}{\textwidth}
\includegraphics[width=0.24\textwidth,clip]{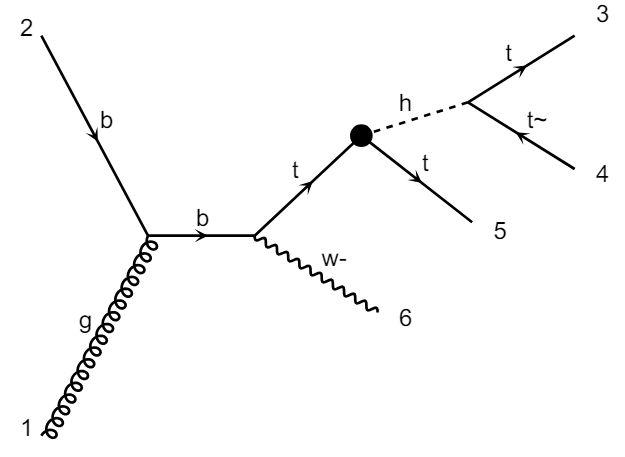}
\includegraphics[width=0.24\textwidth,clip]{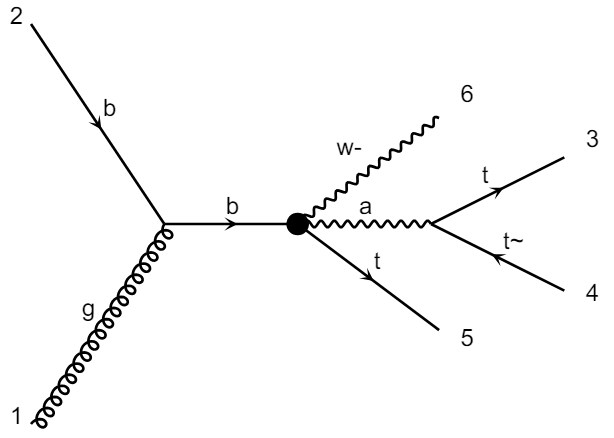}
\includegraphics[width=0.24\textwidth,clip]{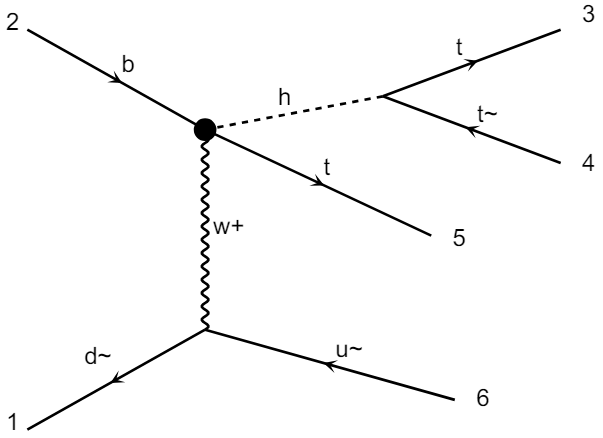}
\includegraphics[width=0.24\textwidth,clip]{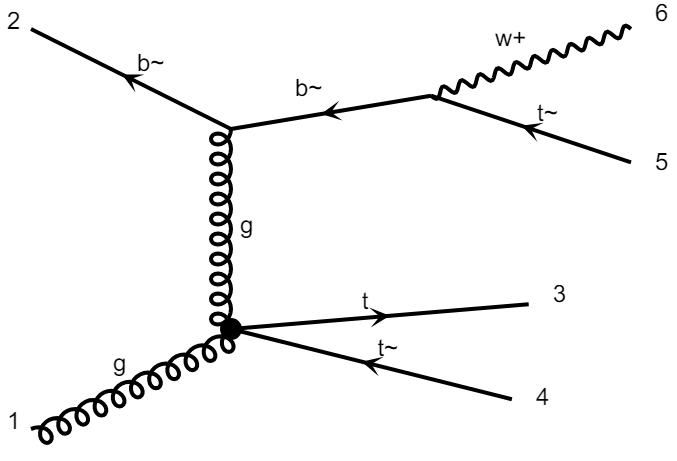}
 \caption{2 heavy fermions and boson fields}
\label{diag_eft:2hb}
 \end{subfigure}

\caption{Representative diagrams for $pp \rightarrow t \bar{t} t(\bar{t}) + X$ with insertion of effective vertices of types: a) 4 heavy fermions, b) 2 heavy and 2 light fermions, c) 2 heavy fermions and boson fields}
\label{3top_diag_eft}
\end{figure}

Despite these challenges, three-top production remains a well-motivated target for BSM studies. First, several BSM scenarios can enhance the three top-quark production rates \cite{Chen:2014ewl, Cao_2019}, including models with flavor-changing neutral currents (FCNC) and top-philic new resonances. Second, compared to $t\bar{t}t\bar{t}$, three top-quark production can probe a broader set of SMEFT directions at leading EFT order. The full set of relevant SMEFT operators is provided in Table \ref{oper_def}. Representative diagrams with insertions of effective vertices are also provided in Figure \ref{3top_diag_eft}.

A comparison of three and four top-quark production processes in terms of their sensitivity to four-heavy SMEFT operators have also been done \cite{Aleshko:2023rkv}. Projections of theoretical constraints on Wilson Coefficients of four-heavy fermion SMEFT operators for both channels are shown in Figure \ref{34t_comparison} \textbf{(b)}. A representative finding is that the three top-quark production process is rather sensitive and can potentially provide complementary constraints on operators with interaction of left fermion currents: $C_{QQ}^1$ and $C_{QQ}^8$. As for operators with right top-currents, while providing similar or worse limits, the three-top data is still potentially useful in the context of global analyses, as can be seen from combined fits to both channels.

\begin{figure}[t]
\centering
\begin{subfigure}[tp]{0.48\textwidth}
    \centering
    \includegraphics[width=0.98\textwidth,clip]{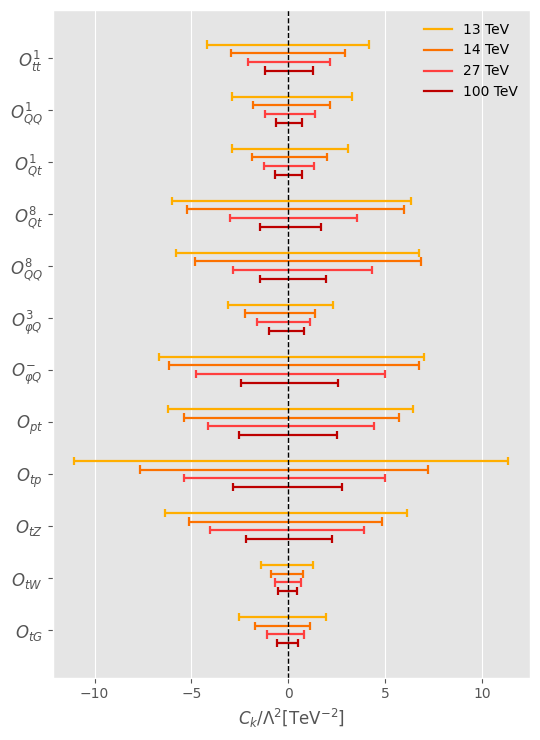}
    \caption{ }
\end{subfigure}
\begin{subfigure}[tp]{0.48\textwidth}
    \centering
    \includegraphics[width=0.98\textwidth,clip]{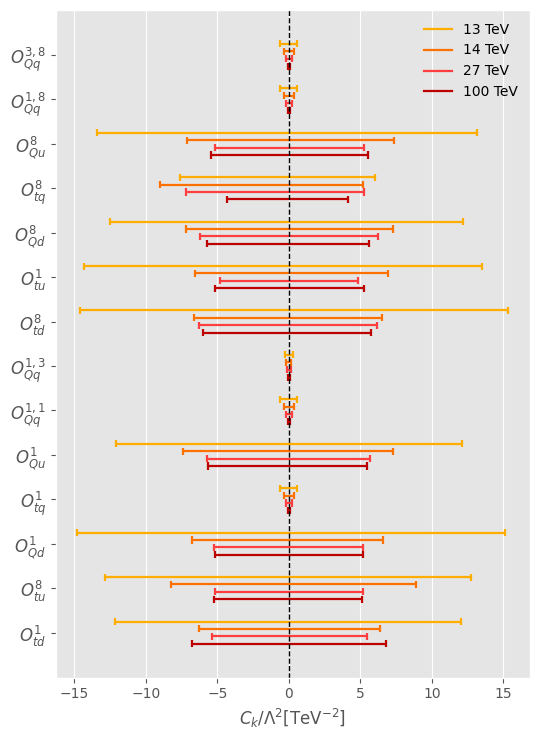}
    \caption{ }
\end{subfigure}
\vspace{-2mm}
\caption{Projections of expected limits on $c_i/\Lambda^2$[TeV$^{-2}$] of SMEFT operators of classes 4-heavy, 2-heavy-boson (a) and 2-heavy-2-light (b) obtainable from three top-quark production process~\cite{Aleshko:2025jua}. Considered scenarios are: LHC (13 TeV, 138 fb$^{-1}$), HL-LHC (14 TeV, 3 ab$^{-1}$), HE-LHC (27 TeV, 15 ab$^{-1}$ and FCC (100 TeV, 25 ab$^{-1}$).}
% \label{smeft-dim6-4top}
\label{3t_limits_proj}
\end{figure}

Potential sensitivity to a full set of top-related dimension six SMEFT operators was also assessed, and corresponding limits on Wilson Coefficients were obtained \cite{Aleshko:2025jua}. Projections to a several high-energy scenarios of bounds on Wilson Coefficient for all relevant classes of operators are provided in Figure \ref{3t_limits_proj}. The first important conclusion drawn from these results is that three top-quark production exhibits relatively strong sensitivity to operator structures involving left-handed light-quark currents ($O_{Qq}^{1-3,1-8}$, $O_{tq}^{1}$), as well as to the dipole operators $O_{tW}$ and $O_{tG}$. A second observation is that the relative hierarchy of "sensitivities" across these directions is stable under projections to higher-energy collider scenarios. Finally, a tangible improvement is expected when moving from 13 to 14 TeV, due to a substantial decrease in the dominant statistical uncertainty component. However, a further increase in both energy and luminosity is expected to provide a modest improvement for the very same reasons as for the $t \bar{t} t\bar{t}$ production: theoretical uncertainty starts to dominate over experimental one.

Summarizing the above, the three top-quark production process is a potentially powerful target for BSM searches; however, there are some limiting factors at the moment. First, similarly to four top-quark production, there are relatively large theoretical uncertainties, and more robust predictions for both SM and SMEFT cases are required. Second, there is the very small SM rate and the partial degeneracy with four top-quark production in multi-lepton final states. The latter requires either better dedicated discriminants to accurately separate the processes, or inclusion of both channels to simultaneous fits. In both cases, better predictions for SM baseline and SMEFT contributions are required.

\section{Unitarity constraints}
\label{sec:unitary_contraints}

One of the potential problems arising in SMEFT is the energy growth of amplitudes induced by higher-dimensional operators. This fact reflects that the EFT is a low energy approximation and must ultimately be UV-completed. The practical concern is that the truncated EFT prediction can violate perturbative (partial-wave) unitarity, signaling the breakdown of the expansion \ref{eft_lagr}. Therefore, to obtain consistent results within the EFT framework, one must quantify such effects and impose corresponding restrictions on an analysis.

Perturbative unitarity bounds are most commonly derived from a partial-wave decomposition \cite{JACOB1959404} of $2\rightarrow2$ scattering amplitudes. Schematically, one can decompose a matrix element into partial waves and exploit $S$-matrix unitarity condition $S^\dagger S = 1$ to obtain constraints on the partial-wave coefficients $a_J$ of the form:
\begin{equation}
	|\mathrm{Re}(a_J)| < \frac{1}{2}
\end{equation}

\noindent These bounds are process dependent: the relevant $a_J$ must be evaluated for the specific partonic subprocesses and helicity configurations. For a detailed discussion of the formalism, we refer to the literature \cite{rauch_2016, Corbett_2017, Maltoni_2019, Cohen_2022, cao_2024}. However, computing the full multi-body, multi-helicity structure is too cumbersome in many cases when the goal is to estimate an EFT validity domain. In collider applications, a pragmatic and typically conservative approach is to identify the dominant energy-growing subamplitudes (often effectively $2\rightarrow2$ configurations embedded in a more complex final state), derive the corresponding unitarity limit as a maximum partonic energy, and implement it as a validity cut on an accessible energy proxy (e.g. an invariant mass) in simulations and fits.

\begin{figure}[t]
	\centering
	\includegraphics[width=0.4\textwidth,clip]{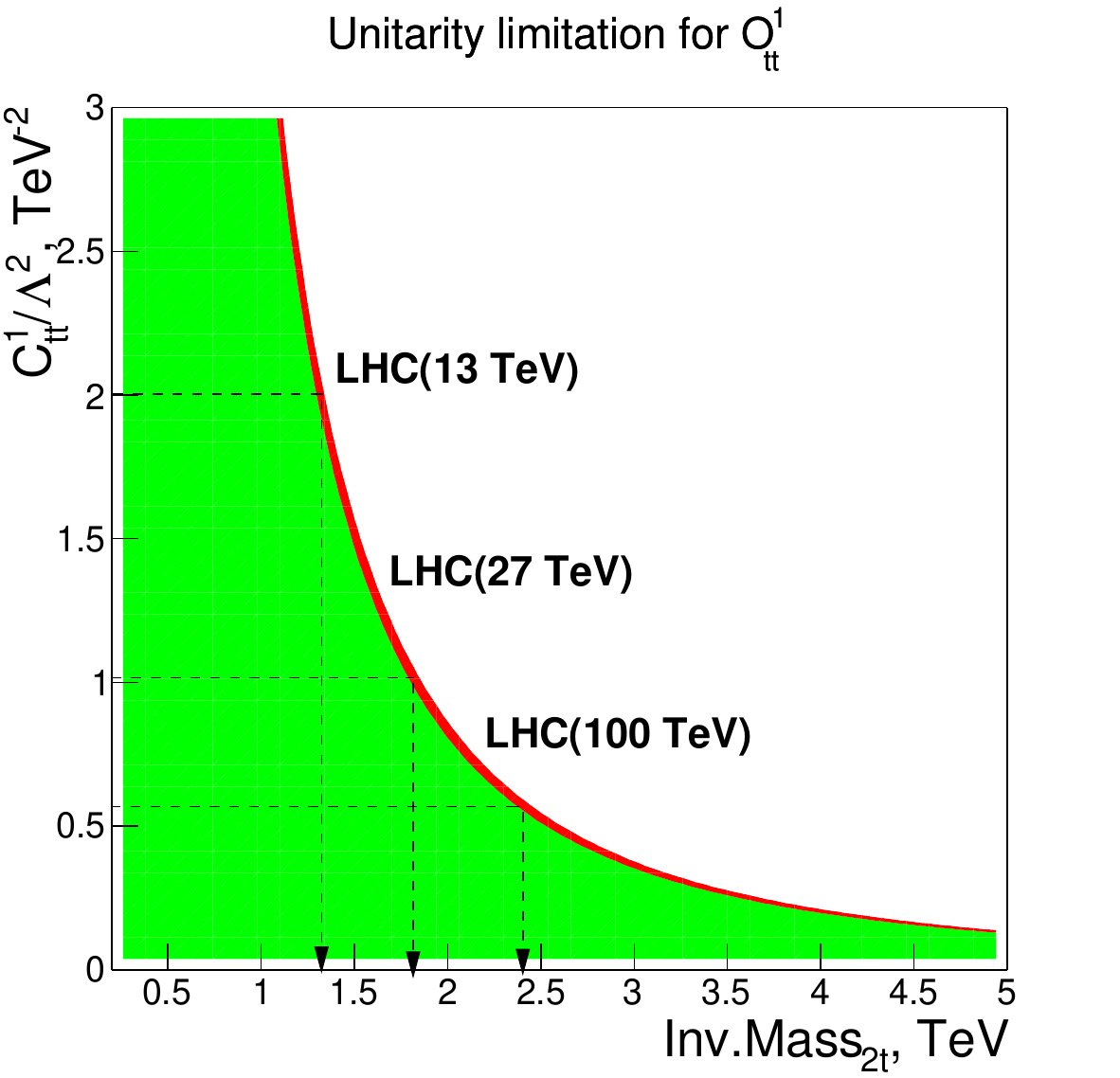}
	\includegraphics[width=0.4\textwidth,clip]{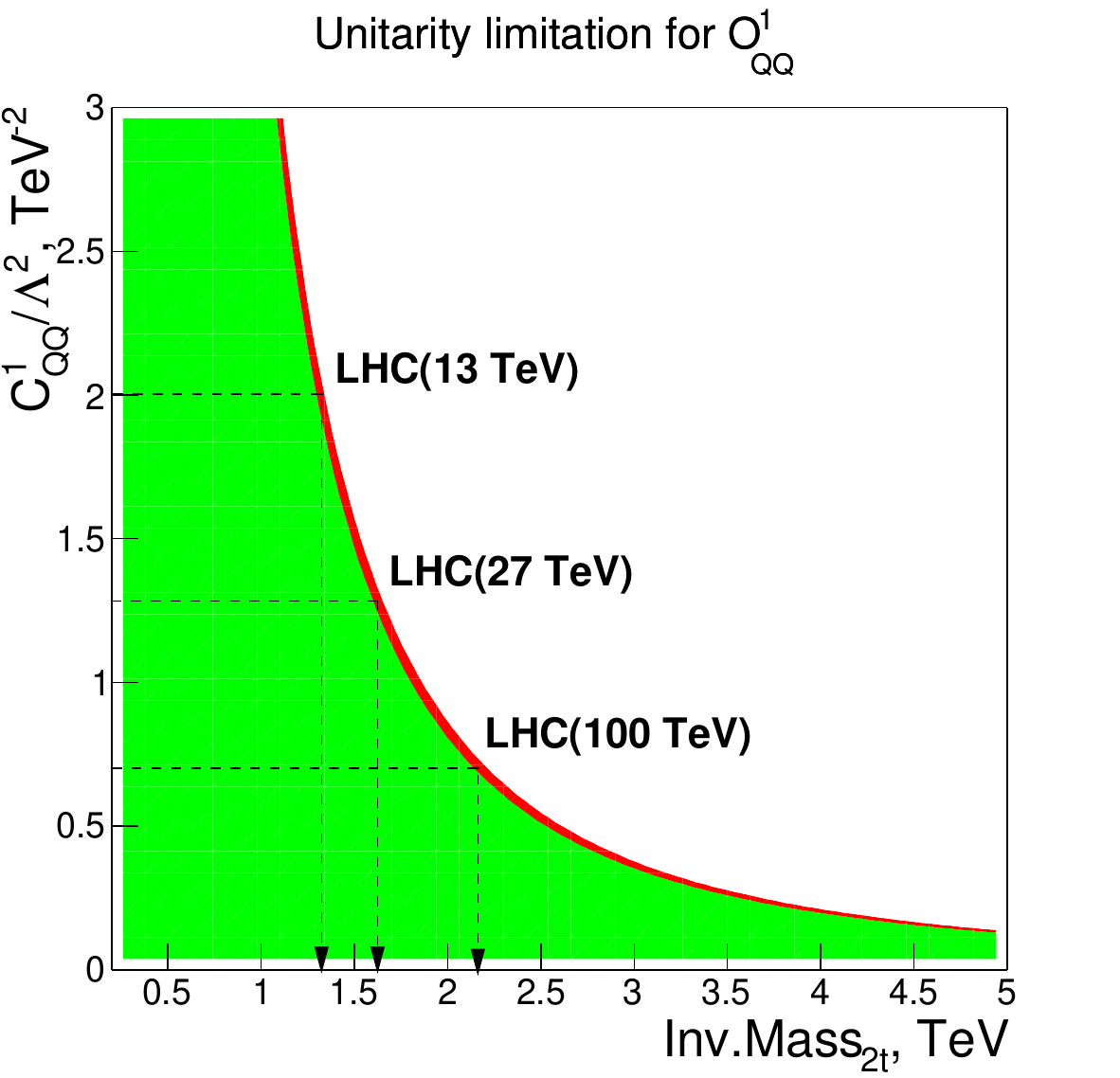}
	\caption{Perturbative unitarity limit (red line) for $O^{1}_{tt}$ (left) and $O^{1}_{QQ}$ (right). The green zone corresponds to the allowed area. The dashed horizontal line indicates the experimental limits on the corresponding Wilson coefficients for different LHC energy values. Dashed vertical lines with an arrow indicate the limits of the invariant mass of two t-quarks for corresponding LHC energy values \cite{Aleshko:2023rkv}}
	\label{fig_limit}      
\end{figure}

As an explicit example in the four-fermion SMEFT sector, in \cite{Aleshko:2023rkv} analytic partial-wave amplitudes for the subprocesses $tt\to tt$ and $t\bar{t}\to t\bar{t}$ were derived, which constitute the leading energy growth induced by the operators $O^1_{tt}$, $O^1_{QQ}$, $O^1_{Qt}$, $O^8_{Qt}$ and $O^8_{QQ}$. These expressions allow one to compute the corresponding perturbative unitarity boundaries and to represent them as limits on the kinematic scale, for instance, in terms of the top-pair invariant mass in the case considered. In Figure \ref{fig_limit} graphs of the perturbative unitarity bounds are drawn in terms of invariant mass of a pair of top quarks and the current upper experimental limits for WC for operators $O^1_{tt}$ and $O^1_{QQ}$. Plots for other operators have a similar structure.

In that setup, the inferred perturbative-unitarity boundary for LHC conditions ($\sqrt{s} = 13$ TeV) corresponds to $m_{tt}\sim1.5$ TeV, while the analogous estimates for scenarios $\sqrt{s} = 27$ TeV and $\sqrt{s} = 100$ TeV increase only moderately, to $\sim2$ TeV and $\sim3$ TeV, respectively. This situation is due to the fact that the cross sections for processes with four top quarks are accumulated at invariant masses not exceeding 10 TeV. A key implication is that, for considered 4-fermion operators, perturbative unitarity can already restrict the use of the truncated SMEFT description at $m_{tt}\sim1-3$ TeV. By comparison, unitarity-based validity estimates reported for anomalous $Wtb$ structures in single-top production \cite{Boos:2023slg} or for $tqg$ FCNC operators \cite{Boos:2020kqq} are typically at substantially higher kinematic scales - on the order of 10 TeV and above.

Finally, the numerical impact of imposing unitarity-motivated validity cuts on the extracted limits on Wilson Coefficients was also assessed. For the 13 TeV scenario, applying such cuts degrades the projected sensitivity to the relevant Wilson coefficients by roughly 5-10\%. While moderate in size, the fact that this effect appears already under current LHC conditions motivates a more systematic treatment of EFT validity in multi-top SMEFT fits.

\section{Summary}

In this review we have summarized recent progress in constraining relevant dimension-six SMEFT Wilson coefficients using three and four top-quark production processes. We also provided a brief introduction to the SMEFT framework and discussed a major limitation of collider EFT interpretations, including practical criteria based on perturbative (partial-wave) unitarity.

One of the main problems in the rare multi-top program is whether three and four top-quark production processes can be experimentally disentangled. Given the strong overlap of $pp \rightarrow t\bar{t}t\bar{t}$ and $pp \rightarrow t\bar{t}tX$ signatures in current multi-lepton selections and the large anti-correlations observed, it seems that a pragmatic near-term approach for SMEFT analyses is a simultaneous interpretation (joint fits) of $pp \rightarrow t\bar{t}t\bar{t}$ and $pp \rightarrow t\bar{t}t$ contributions. 

Should dedicated $pp \rightarrow t\bar{t}t$ measurements become feasible with sufficient separation power, three top-quark production may provide complementary constraints on some of WC directions. In particular, existing studies indicate that three top-quark production can be sensitive to four-fermion structures involving left-handed quark currents, including operators built from two left-handed heavy-quark currents and from one left-handed light-quark current contracted with a heavy-quark current.

Other limiting factors are also clear: reduced experimental uncertainties, higher-precision SM baselines and SMEFT predictions, and a consistent treatment of EFT truncation and validity are required for further progress in multi-top SMEFT constraints.

\bibliographystyle{unsrt}
\bibliography{smeft_34tops}

\end{document}